# Spaced TiO$_2$ Nanotube Arrays Allow for High Performance Hierarchical Supercapacitor Structure

Nhat Truong Nguyen,[a,†] Selda Ozkan,[a,†] Imgon Hwang,[a] Xuemei Zhou,[a] Patrik Schmuki[a,b,*]

[a]Department of Materials Science and Engineering WW4-LKO, University of Erlangen-Nuremberg, Martensstrasse 7, D-91058 Erlangen, Germany.

[b]Chemistry Department, Faculty of Sciences, King Abdulaziz University, 80203 Jeddah, Saudi Arabia Kingdom

[†] These authors contributed equally to this work.

*Corresponding author. Email: schmuki@ww.uni-erlangen.de

Link to the published article:

https://pubs.rsc.org/en/content/articlehtml/2017/ta/c6ta10179h

# Abstract

In this work we describe the synthesis and electrochemical properties of nitridated hierarchical $TiO_2$ nanotubes as an electrode for supercapacitors. The hierarchical $TiO_2$ nanostructures are formed by a controlled layer-by-layer $TiO_2$ nanoparticle decoration on self-organized spaced $TiO_2$ nanotubes. These structures are then annealed in $NH_3$ atmosphere at elevated temperature to convert the material to a nitride structure – this drastically enhances their electron-transport properties. The areal capacitance of hierarchical structures can be tuned by changing the number of decorated $TiO_2$ nanoparticle layers. The capacitance enhancement of the hierarchical structures reaches a maximum when the surface area through nanoparticle deposition is highest and the conductivity via nitridation is optimized.

Due to the increasing energy demands, energy conversion and storage technologies have been intensively developed in recent years. Supercapacitors are considered a very promising candidate for electrochemical energy storage due to their high power density, long cycle life and fast charge-discharge rate.[1–3] According to charge storage mechanism, supercapacitors are generally classified into two categories: electrical double layer capacitors (EDLCs) and pseudocapacitors.[4,5] EDLCs are based on non-faradic charge separation at the electrode/electrolyte interface while pseudocapacitors are dominated by a faradaic process at the electrode – this allows pseudocapacitors to provide a higher specific capacitance and energy density via the charging/discharging of electrode internal contribution of redox-states.[4]

Among a wide range of materials examined as supercapacitor electrodes, hydrous $RuO_2$ remains the champion material due to its fast and reversible intercalation properties, and a high electron conductivity.[6,7] However, the high cost of $RuO_2$ and increasing demand for supercapacitors have initiated extensive research for alternative materials. Particularly transition metal oxides such as NiO, $V_2O_5$, $MoO_3$, $MnO_2$, $Co_3O_4$ and $TiO_2$ have been considered as potential electrode materials.[8–13] Among them, $TiO_2$ is one of the most interesting materials for energy storage since it is abundant, low cost and chemically stable.[14,15]

In order to provide an ideal geometry for charge storage, the most advantageous electrode is one that offers densely dispersed nanoparticles of a few nanometers (corresponding to the diffusion length of the intercalating species) and a geometrically optimized electron-conductive scaffold. To address these prerequisites, one-dimensional $TiO_2$ structures, *i.e.*, nanotubes (NTs) that possess directional charge transport and a high surface area have been increasingly explored for their supercapacitor performance.[16,17] The most straightforward method to generate a suitable $TiO_2$ nanotube electrode is by self-organizing anodization of Ti in a suitable electrolyte.[15,18] The resulting $TiO_2$ NTs are vertically aligned to the substrate, their diameter and length can be easily adjusted, and their attachment to the Ti metal substrate provides an ideal electrode configuration.[19]

In general, to grow self-organized NTs aqueous, glycerol or ethylene glycol-based electrolyte are used which results in hexagonally close-packed arrays of $TiO_2$ NTs.[15] This arrangement limits the use of these $TiO_2$ NT layers as a scaffold for deposition of nanoparticles, *i.e.*, to increase their electrochemical performance by constructing hierarchical assemblies. Furthermore, typically the supercapacitive performance of 3D-$TiO_2$ structures is strongly hampered due to their low electronic conductivity. Therefore, not only optimizing the $TiO_2$ nanostructure but also developing measures to increase the intrinsic $TiO_2$ conductivity are needed.

In the present work, we show that nitridation of hierarchical $TiO_2$ NTs based on spaced tubes through a heat treatment process under $NH_3$ atmosphere at elevated temperature can yield highly beneficial effects. We demonstrate that an adequate $NH_3$ treatment leads to nitrogen stabilized $Ti^{3+}$ states that provide a strongly enhanced conductivity that in turn leads to a significant enhancement in electrochemical performance.

The spaced $TiO_2$ NTs used in this work (**Figure 1**) were fabricated by anodization of titanium substrates in a triethylene glycol (TEG) solution (see SI). The as-formed $TiO_2$ NTs have a diameter of 200 nm and a length of 7 µm (Figure 1a and S1). The main difference to conventional $TiO_2$ NTs (Figure S2a) is that the tubes shown in Figure 1 provide a defined and controllable intertube spacing that allows the deposition of a secondary material in the free space between the tubes. This spacing in our case was adjusted to 200 nm. To increase the surface area of NTs, a defined conformal layer of $TiO_2$ nanoparticles were decorated on the $TiO_2$ NTs by a $TiCl_4$-hydrolysis approach.[20,21] These nanoparticles have a diameter of 5-8 nm and with an increasing number of $TiCl_4$ treatments the inner and outer tube wall become increasingly decorated (Figure 1b), thus steadily filling space between the tubes. On contrary, hexagonally close-packed $TiO_2$ NTs show a completely blocked entrance after just a few decorated layers (Figure S2b, c). After decoration the samples were thermally treated in $NH_3$ atmosphere at different temperatures (400°C to 700°C) for 1 h to induce nitridation. The color of the samples turned from dark blue to dark green or black depending on the temperature of the $NH_3$ treatment (Figure S3). Furthermore, clearly the nanoparticle size increases with an increasing $NH_3$ treatment temperature, *i.e.*, sintering occurs, the more drastic the higher the treatment temperature (Figure S4). In other words while the decoration strongly increases the surface area, the $NH_3$ treatment at elevated temperature reduces this again, the higher the temperature the more significant. This is confirmed by dye loading measurements (Figure S5 and S6), that reflects the specific surface area.[22] Clearly, nanoparticle decoration strongly increases the specific surface area up to 6 times (from dye loading of 22 to 138 nmol cm$^{-2}$) while $NH_3$ treatment at 700°C decreases their surface area by a factor of 2 (61 nmol cm$^{-2}$). However the $NH_3$ treatment with increasing temperature leads to a drastic enhancement of conductivity of the material (Figure 1c, d).[23–25] Solid state conductivity measurements show the conductivity of the hierarchical structures to increase by a factor of 400 after $NH_3$ treatment at 400°C and a factor of $10^7$ at 600°C. Treatments at 700°C do not significant enhance the conductivity further.

Figure 1e and f show the drastic effect of the combined nanoparticle decoration and nitridation on the capacitive electrochemical response of the $TiO_2$-hierachical structure-$NH_3$ electrodes (S-Tx-$NH_3$, x is the number of $TiCl_4$ treatment). Cyclic voltammetry (CV) tests were conducted in a three-electrode electrochemical cell with a Pt sheet counter electrode and a Ag/AgCl reference electrode in 0.5 M aqueous $Na_2SO_4$ solution at a scan rate of 100 mV s$^{-1}$. The CV curves show a rectangular shape thus indicating the EDLCs and the capacitcance of the electrode can be obtained by equation (1) in the SI. Bare $TiO_2$ NTs obviously show a much smaller current density and thus a capacitance of 11.4 mF cm$^{-2}$ (at a scan rate of 100 mV s$^{-1}$) in comparison to the hierarchical structures (2 layers of $TiO_2$ nanoparticles) where 2.3 times higher current density and a capacitance of 26.6 mF cm$^{-2}$ are reached (Figure 1e). With an increasing number of decorated layers, the capacitive performance increases from 11.4 to 42.4 mF cm$^{-2}$ (up to 4 times) and is maximized after eight layers of nanoparticles. For higher amounts of nanoparticle loading, the current density decreases since the structures are completely filled with nanoparticles thus limiting the particle/electrolyte interface (Figure S7). The effect of nitridation temperature on electrochemical behavior is apparent from Figure 1f. The capacitive current density of $TiO_2$ samples increases by a factor of 50 with the increase of annealing temperature up to 600°C. A decrease of current density was obtained for the samples annealed in $NH_3$ at 700°C. This drop of current can be ascribed to the coarsening of nanoparticles which reduces the available area for intercalation on the structure.

**Figure 2a** shows the X-ray diffraction (XRD) patterns of the $TiO_2$ samples annealed in air at 450°C and in $NH_3$ at different temperatures. For $TiO_2$ NTs and hierarchical $TiO_2$ structures (after 8 times $TiCl_4$ treatment-denoted as S-T8) the XRD peaks can be ascribed to anatase and a small amount of rutile phase (peak at $2\theta = 25.3°$ and $2\theta = 27.4°$, respectively).[26–28] After nitridation at 400-600°C, the XRD-spectra do not change. However at 700°C, no anatase peaks can be detected anymore but peaks indicating the formation of titanium nitride, as a cubic phase, become visible at $2\theta = 42.6°$.[29–31] This is in line with literature that shows annealing at 800°C can transform $TiO_2$ into TiN.[29,31]

To examine the effect of $NH_3$ treatments on chemical composition of hierarchical $TiO_2$ structures, we performed X-ray photoelectron spectroscopy (XPS) (Figure 2b-d and S8). Figure 2d and S8 show the Ti2p core level XPS spectra of $TiO_2$ and $NH_3$-treated $TiO_2$ samples. Clearly the Ti2p peaks for $NH_3$ treatment above 400°C appears with a typical $Ti^{3+}$ tail while the Ti2p spectrum for the treatment at 400°C remains unchanged.[32] The Ti2p peak can be deconvoluted into three peaks. The doublet peak at 458.9 eV for $Ti2p_{3/2}$ and 464.6 eV for $Ti2p_{1/2}$ are assigned to $Ti^{4+}$. The $Ti^{3+}$ in a Ti-N configuration can be assigned to the second doublet of the Ti2p peak since the N is mainly introduced in a substitutional position (Figure 2c).[33,34] The third doublet of the Ti2p peak (455.9 eV for $Ti2p_{3/2}$ and 461.7 eV for $Ti2p_{1/2}$) is assigned to $Ti^{2+x}$, in accord with literature,[35–37] and as indicated by the shift of the O1s peak to a higher valence state and the decrease of the peak intensity at 700°C (Figure 2b). The deconvolution results for all samples are listed in Figure 2e and clearly TiO$_2$ sample annealed in NH$_3$ at 700°C has the largest amount of Ti$^{3+}$ (36.86 %) in comparison to sample annealed at 600°C and 500°C (14.30 and 10.77 %, respectively). And the treatment at 700°C leads to even lower Ti$^{2+x}$ states with a ratio of 12.36%. For the treatment at 400°C and the non-NH$_3$ treated samples, only the presence of Ti$^{4+}$ was detected. This indicates the formation of titanium nitride for samples annealed in NH$_3$ which can be explained that at temperature above 550°C, NH$_3$ decomposes forming H$_2$ thus introduces a strong reducing atmosphere – that can reduce Ti$^{4+}$ to lower valence states.[38,39] This is also in line with high resolution O1s spectra where the intensity of peaks at 530.2 eV decreases significantly compared to treatment at lower temperatures (Figure 2b).[40] A small shift in O1s peak for 700°C sample is ascribed to the formation of O-N bonding.[41] In line, N1s spectra (Figure 2c) show clear peaks for nitrogen (at 396.5 eV) appearing in the NH$_3$-treated samples at 500°C, 600°C and 700°C while there are no such peaks for untreated TiO$_2$ sample and treated sample at 400°C. This peak is usually ascribed to substitutional nitrogen in TiO$_2$ thus can indicate the successful nitridation of hierarchical TiO$_2$ nanostructures at the temperature above 400°C.[29] The dramatically increased intensity of N1s peaks for samples annealed at 700°C in NH$_3$ demonstrates that large amount of nitrogen has been introduced to TiO$_2$ and forms titanium nitride that is in accordance with XRD data. Please also note that the nitrogen peak shifts towards a higher binding energy (397.3 eV) that indicates, except for titanium nitride formation, additional substitutional nitrogen exists in the lattice in a Ti-O-N configuration. This is in line with a clear shift of O1s peak to higher binding energy (oxygen vacancy formation) for sample treated at 700°C.[37,42,43] Furthermore, the atomic concentration of nitrogen increases with an increasing NH$_3$ temperature (from 0.94 to 28.33 at%) while the oxygen concentration decreases by a factor of 2 (from 67.66 at% at 400°C to 36.84 at% at 700°C). The above XPS results confirm that the nitrogen concentration in TiO$_2$ is strongly correlated to the annealing temperature in NH$_3$.[24]

Regarding the electrochemical performance, the $TiO_2$ samples nitridated at 600°C yields the highest areal capacitance of 85.7 mF cm$^{-2}$ at a scan rate of 10 mV s$^{-1}$ (**Figure 3a**). The charge-discharge curves in Figure 3b further confirm that an optimized condition for $NH_3$ treatment is established at 600°C. Furthermore, all $NH_3$-treated samples have a very smaller IR drop compared with non-treated samples (0.02 and 0.18 V, respectively – Figure S9), proving a low internal resistance for the nitridation $TiO_2$ NTs (in line with conductivity measurements).

Figure 3c exhibits the highest values for the nitridated hierarchical structure at 600°C at a current density up to 3 mA cm$^{-2}$. The difference in capacitance obtained from CV curves and charge-discharge curves is due to the selection range of sweep rate and charge-discharge current density. Figure 3d shows the Mott-Schottky plots based on capacitance that were acquired from the electrochemical impedance at 100 Hz. The $TiO_2$ samples up to 500°C clearly exhibit a characteristic of a n-type semiconductor. The carrier densities of the $NH_3$ treated samples at 400°C and 500°C are $1.6 \times 10^{23}$ and $1.0 \times 10^{24}$ cm$^{-3}$ while that for non-treated sample is $5.2 \times 10^{19}$ cm$^{-3}$ (employing a dielectric constant of $TiO_2$ $\varepsilon \approx 31$).[44] The samples at 600°C and 700°C show a virtually metallic capacitance behavior in line with the resistivity values of a few $\Omega$ in the solid state IV curves of Figure 1d. In other words, annealing in $NH_3$ first significantly enhances the doping densities of $TiO_2$ NTs by several orders of magnitude and at higher temperatures (600°C and 700°C) establishes a near metallic value – this correlates to the observed formation of TiN species which have great properties as an electrode for supercapacitors such as superior electrical conductivity, extreme hardness and remarkable chemical resistance.[45,46]

Furthermore, to obtain more information on the charge-transfer properties, electrochemical impedance spectroscopy (EIS) measurements were carried out. Figure 3e and f show the Nyquist plots and fitted data measured for different samples at open circuit potential. The results indicate that nitridation significantly decreases the electrode charge transfer resistance. Also in this case, clearly 600°C was found to be an optimized temperature for nitridation that provides the lowest charge transfer resistance per provided electrode area (3.0 KΩ).

CV curves of the S-T8-NH$_3$ were measured at different scan rates of 1 to 200 mV s$^{-1}$ and the results are presented in **Figure 4a**. These nitridated samples show quasi-rectangular CV curves and no pseudocapacitive oxidation/reduction peaks, indicating excellent electrical double layer capacitive properties. The calculated areal capacitance at a high scan rate of 200 mV s$^{-1}$ is 42.4 mF cm$^{-2}$, which results in capacitance retention of nearly 50 % compared with the values measured at 10 mV s$^{-1}$. The performance of nitridated hierarchical structure was further tested by galvanostatic charge-discharge measurements (Figure 4b). The charge-discharge curves are nearly symmetric with a linear relation between charge-discharge and time. The hierarchical electrodes provide a specific capacitance of 45.19 mF cm$^{-2}$ at a current density of 0.2 mA cm$^{-2}$. The hierarchical structures show a better performance compared with conventional TiO$_2$ NTs due to their higher surface area (Figure S10). As shown in Figure 4c, the nitridated hierarchical sample also shows a stable long term cycling performance. The XRD measurements after electrochemical tests remain unchanged, indicating that the structures are stable (Figure S11). The capacitance of hierarchical structures obtained in the present work is significantly higher than other reported results using anodic TiO$_2$ NTs including tubes after different reduction treatments such as electrochemically reduced-or-doped TiO$_2$ nanotubes,[47–51] hydrogen treated nanotubes,[52,53] and even carbon-decorated TiO$_2$ nanotube composites.[54–57]

In summary, we introduced the use of spaced TiO$_2$ NTs for a conformal layer-by-layer decoration of TiO$_2$ nanoparticles. These hierarchical structures via NH$_3$ treatment significantly enhance the electrochemical performance of TiO$_2$ NTs. Sample annealed at 600°C in NH$_3$ yields an optimized specific capacitance. The improvement in capacitance can be attributed to the combination of a high surface area and strongly increased conductivity induced by nitride formation in the tubes. Furthermore, the areal capacitance of hierarchical structures can be adjusted by varying the number of TiO$_2$ nanoparticle layers. In addition, we show that spaced

TiO$_2$ NTs can be used to construct high-performance supercapacitors by a controlled decoration of TiO$_2$ nanoparticles that can be replaced by other materials for other energy storage devices.

**Acknowledgements**

The authors would like to acknowledge the ERC, the DFG and the DFG cluster of excellence, EAM for financial support. Dr. Lei Wang is acknowledged for XRD measurements. S. Hejazi is acknowledged for the evaluation of EIS data.

# Figure

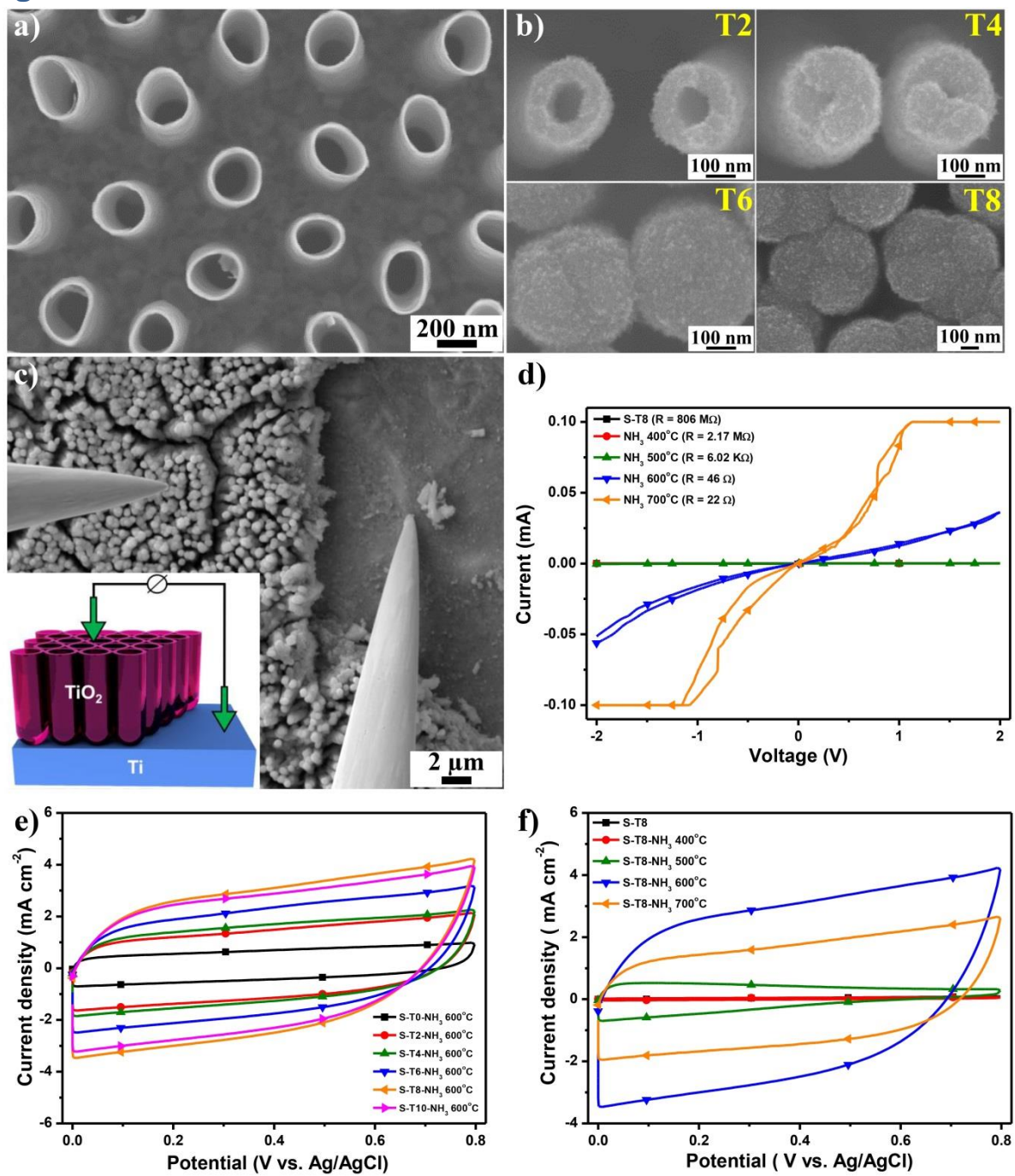

**Figure 1** SEM images of: a) as-formed spaced TiO$_2$ NTs, b) spaced TiO$_2$ nanotubes decorated with different layers of TiO$_2$ nanoparticles (2, 4, 6 and 8 times), c) solid state conductivity measurement with 2 tips, the inset illustrates how the conductivity measurement in the SEM works. d) Conductivity measurements of hierarchical structures without/with NH$_3$ treatment at different temperatures. e) CV curves at a scan rate 100 mV s$^{-1}$ of spaced TiO$_2$ NTs loaded with different layers of TiO$_2$ nanoparticles and annealed in NH$_3$ at 600°C for 1 h. f) CV curves of spaced TiO$_2$ NT decorated with 8 layers of TiO$_2$ nanoparticles and annealed in NH$_3$ at different temperatures for 1 h.

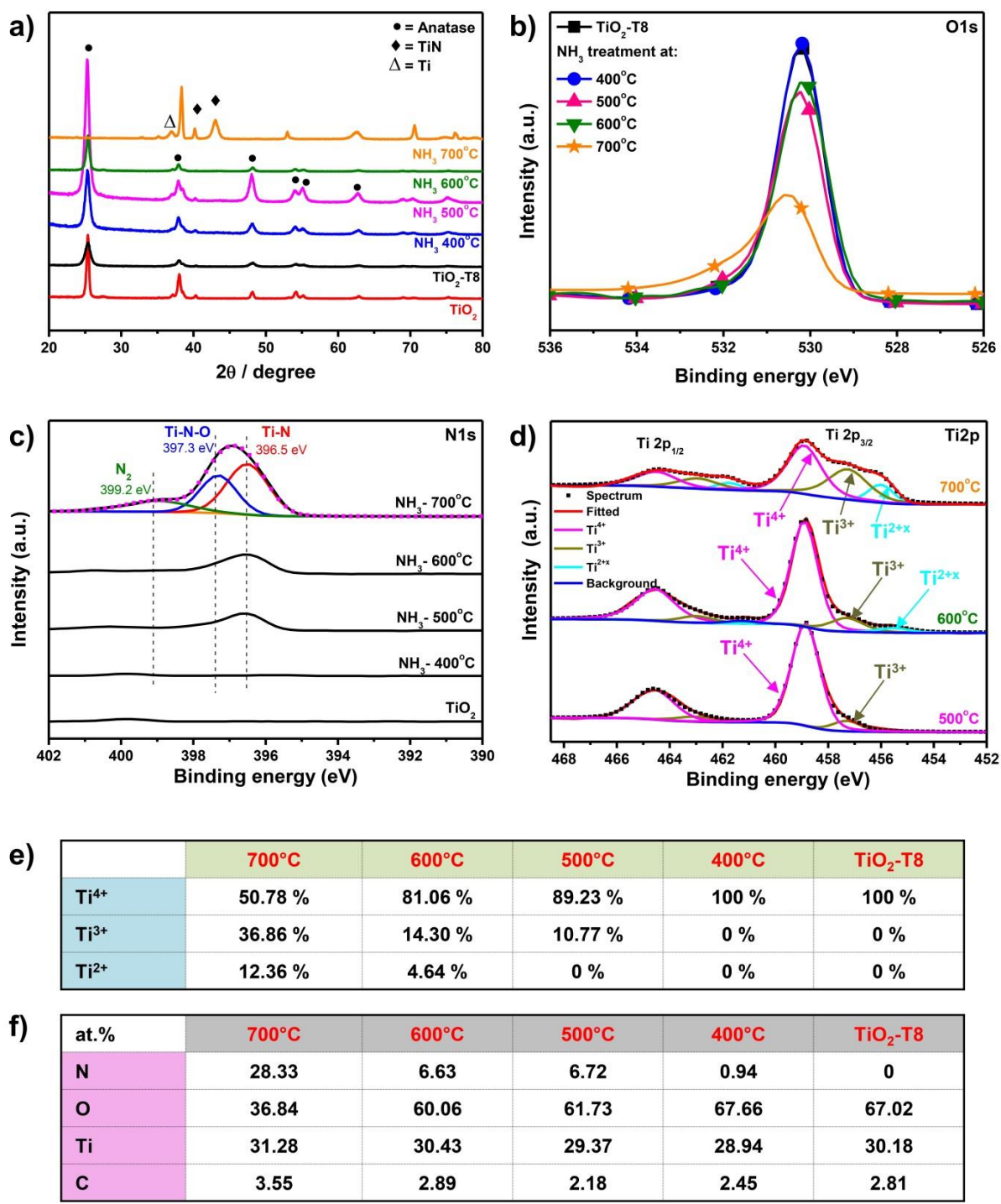

**Figure 2** a) XRD patterns of spaced $TiO_2$ NTs and NTs annealed in $NH_3$ at different temperatures. b) O1s and c) N1s high resolution XPS spectra for the particle-decorated $TiO_2$ NTs and particle-decorated $TiO_2$ NTs annealed in $NH_3$ at 400, 500, 600 and 700°C. d) peak fitting of the Ti2p peaks for the particle-decorated $TiO_2$ NTs annealed in $NH_3$ at 500, 600 and 700°C. e) Calculated concentration of titanium complex and f) atomic concentration for the particle-decorated $TiO_2$ NTs ($TiO_2$-T8) and particle-decorated $TiO_2$ NTs annealed in $NH_3$ at 400, 500, 600 and 700°C.

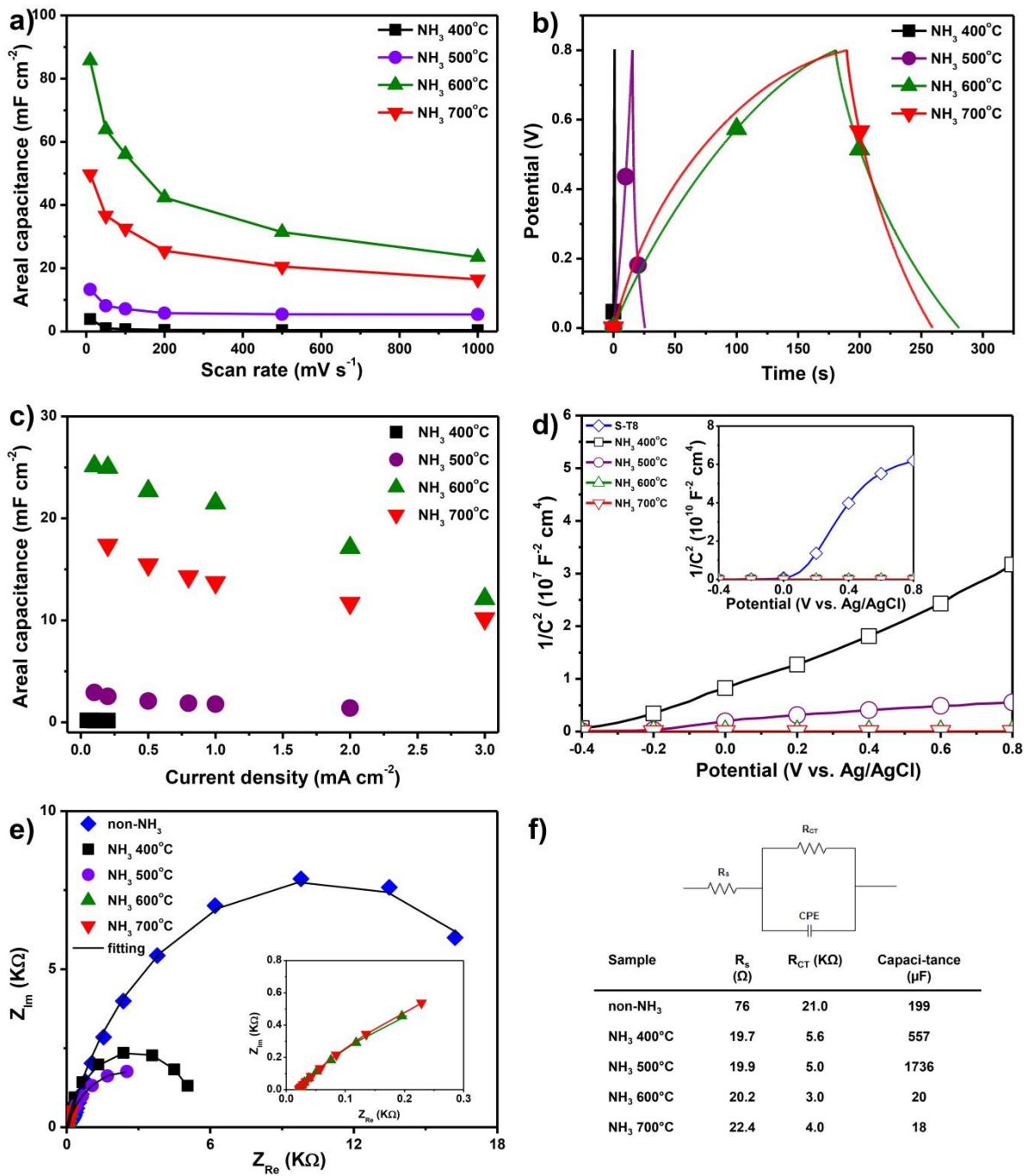

**Figure 3** a) Areal capacitance of TiO$_2$ samples annealed in NH$_3$ at different temperatures measured as a function of scan rate. b) Galvanostatic charge/discharge curves (at a current density of 200 µA cm$^{-2}$) of S-T8 samples annealed in NH$_3$ at different temperatures. c) Areal capacitance of TiO$_2$ samples annealed in NH$_3$ at different temperatures measured as a function of current density. d) Mott-Schottky plots of the TiO$_2$ samples measured at 100 Hz. e) Nyquist plot and f) Equivalent circuit and impedance data for the hierarchical TiO$_2$ NTs with/without NH$_3$ treatment (measured at open circuit potential).
16

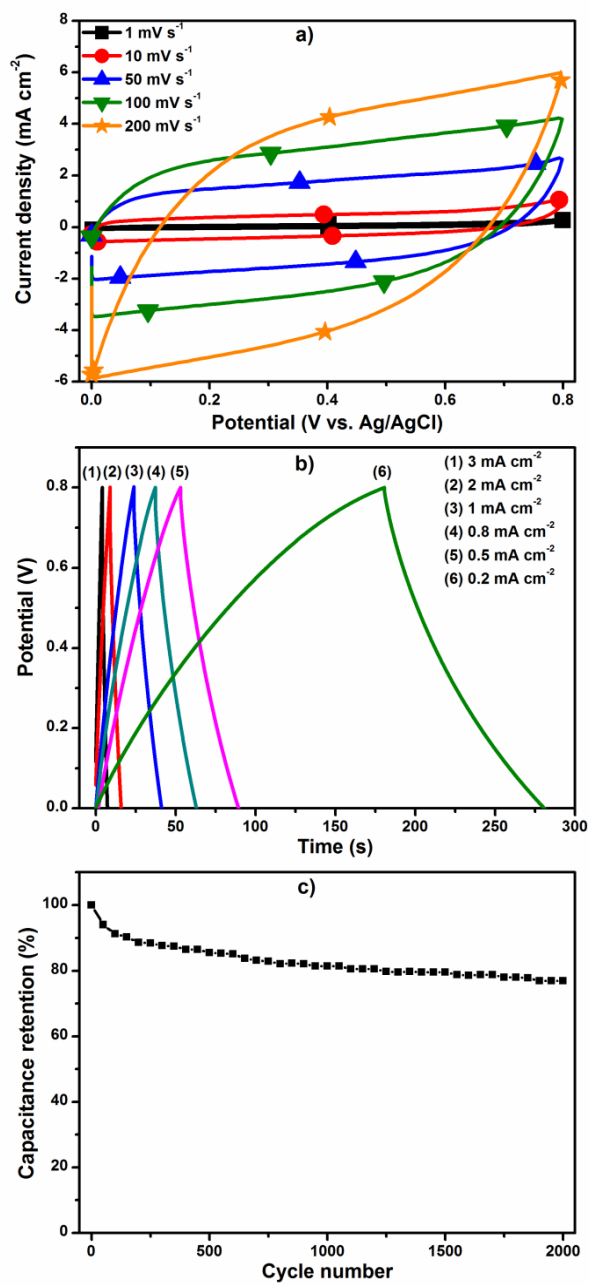

**Figure 4** a) CV curves of S-T8-NH$_3$ at different scan rates. b) Galvanostatic charge/discharge curves of S-T8-NH$_3$ at different current densities. c) Cycle performance of S-T8-NH$_3$ measured at a current density of 3 mA cm$^{-2}$. NH$_3$ treatment was conducted at 600°C for 1 h.